\documentclass[epj]{svjour}
\usepackage{graphics}
\usepackage{graphicx}
\usepackage{latexsym}
\usepackage{hhline}
\usepackage{amssymb}
\usepackage{wasysym}
\usepackage{epsfig}

\newcommand{\be}{\begin{equation}}
\newcommand{\ee}{\end{equation}}
\newcommand{\bea}{\begin{eqnarray}}
\newcommand{\eea}{\end{eqnarray}}
\newcommand{\ba}{\begin{array}}
\newcommand{\ea}{\end{array}}
\newcommand{\bi}{\begin{itemize}}
\newcommand{\ei}{\end{itemize}}
\newcommand{\mi}{\mbox i}

\newcommand{\refe}[1]{(\ref{#1})}

\newcommand{\mck}{{\mathcal K}}
\newcommand{\mcl}{{\mathcal L}}

\newcommand{\mct}{{\mathcal T}}

\newcommand{\ra}{\rightarrow}

\renewcommand{\slash}{/ \!\!\!\!\,}

\newcommand{\foh}{\frac{1}{2}}

\newcommand{\fth}{\frac{3}{2}}
\newcommand{\ffh}{\frac{5}{2}}

\newcommand{\umat}{1 \! \! 1}

\newcommand{\JPR}[3]{Phys. Rev. {\bf #1}, #2 (#3)}
\newcommand{\JPS}[3]{{Phys. Scr.} {\bf #1}, #2 (#3)}
\newcommand{\JPRL}[3]{{Phys. Rev. Lett.} {\bf #1}, #2 (#3)}
\newcommand{\JPL}[3]{{Phys. Lett.} {\bf #1}, #2 (#3)}
\newcommand{\JPRC}[3]{Phys. Rev. C {\bf #1}, #2 (#3)}
\newcommand{\JPRD}[3]{Phys. Rev. D {\bf #1}, #2 (#3)}
\newcommand{\JNP}[3]{Nucl. Phys. {\bf #1}, #2 (#3)}

\newcommand{\JZP}[3]{{Z. Phys.} {\bf #1}, #2 (#3)}

\newcommand{\Jpin}[3]{{$\pi N$-Newsletter} {\bf #1}, #2 (#3)}
\newcommand{\JPPNP}[3]{{Prog. Part. Nucl. Phys.} {\bf #1}, #2 (#3)}
\newcommand{\ibid}[3]{{ibid}, {\bf #1}, #2 (#3)}

\newcommand{\JPRep}[3]{{Phys. Rept.} {\bf #1}, #2 (#3)}

\begin{document}
\title{ Spin-$\ffh$ resonance contributions to the pion-induced reactions \\ for energies 
$\sqrt{s}\leqslant$2.0 GeV.}

\author{V. Shklyar,  G. Penner and U. Mosel      
\thanks{Supported by DFG and GSI Darmstadt  }%
}                     
\offprints{}          
\institute{Institut f\"ur Theoretische Physik, Universit\"at Giessen, D-35392 Giessen, Germany}
\date{Received: date / Revised version: date}
%
\abstract{
 The spin-$\ffh$ resonance effects are studied within the coupled channel 
effective Lagrangian model for baryon resonance analysis. We extend our previous 
hadronic calculations to incorporate  the $D_{15}$,  $F_{15}$,  $D_{35}$,  
$F_{35}$ states.
 While the effect of the spin-$\ffh$ resonances to the
$\eta N$, $K \Lambda$, and $K \Sigma$ reactions are small, the contribution 
to  the $\omega N$ is found to be important. The results for the 'conventional' and
Pascalutsa-like spin-$\ffh$ descriptions are discussed. 
}

\authorrunning{V. Shklyar  et al.}

\titlerunning{Spin-$\ffh$ resonance...}

\maketitle

\section{Introduction}
\label{intro}
The extraction of  baryon-resonance properties is one of the important 
tasks of modern hadron physics. Great efforts have been made in the past to 
obtain this information from the analysis of pion- and photon-induced  reaction data. 
The precise knowledge of these properties is an important step towards understanding 
the hadron structure and finally the strong interactions. 

Some quark models (see \cite{Capstick02} 
and Refs. therein) predict  
that the  baryon resonance spectrum may be richer then discovered  so far. 
This is the so-called problem of 'missing' nucleon resonances. One assumes that these states
are weakly coupled to  pion channels and 
are consequently not clearly seen in  $\pi N$, $2\pi N$ and $\eta N$ reactions from 
which experimental
data are most often used for baryon-resonance analyses. 
To incorporate other possible finale states a unitary coupled-channel 
model (Giessen model) has been developed which includes  $\gamma N$, $\pi N$, $2\pi N$, 
$\eta N$, $K \Lambda$  final states and deals with all available experimental
data on pion- and  photon-induced reactions \cite{feusti98,feusti99}. The most recent
 extensions
of this model include $K \Sigma$ and $\omega N$ final states \cite{PennerPRC65,Penner02,pm2} as well, 
which allows for the simultaneous analysis of all hadronic and photoproduction data 
up to $\sqrt s =$ 2 GeV. 
A shortcoming of this study is the missing of higher-spin resonances with spin 
$J> \fth$.  Since the \mbox{spin-$\ffh$}  resonances have large electromagnetic couplings 
\cite{pdg,arndt02,maid} 
this limited the previous analysis of the Compton scattering data to the energy region 
$\sqrt{s}\leqslant$ 1.6 GeV.
Moreover, the extension to higher-spin baryon spectra becomes unavoidable for investigation of
'hidden' or  'missing' nucleon  resonances. In particular, a  study  of the
spin-$\ffh$ part of the baryon spectra can shed light on the dynamics of the vector
($\omega$ and $\rho$) meson production mechanisms
which is itself a  very intriguing question (see \cite{Titov}
and references therein). 

In the present paper we study the effect of spin-$\ffh$ resonance contributions to 
$\pi N$, $2\pi N$, $\eta N$, $K \Lambda$, $K \Sigma$, and $\omega N$ final states. Starting  from  
the effective Lagrangian coupled-channel model 
\cite{Penner02}  we extend our previous hadronic calculation \cite{Penner02} by including 
the $D_{15}$, $F_{15}$, $D_{35}$, $F_{35}$ resonances and simultaneously analysing all
available pion-induced reaction data in up to 2 GeV energy region. Due to the 
coupled-channel calculations 
this model provides a
stringent test for the resonance contributions to the all open  final states. 
Similar to the spin-$\fth$ case in \cite{Penner02,pm2}, the contributions from  
spin-$\ffh$ states are investigated for two different types of the spin-$\ffh$ couplings: 
for the 'conventional' ($C$) and Pascalutsa ($P$) prescriptions. While the first approach dates back to the
original work of Rarita and Schwinger \cite{rarita} and is widely used in the literature, the 
latter one assumes the gauge-invariant resonance coupling.  Although the data quality is not good enough
to distinguish between these two pictures now, this question is challenging for  an understanding of
the meson-baryon interactions. With this aim in mind,  the present work extends
our  earlier   
multi-channel analysis based on an  effective Lagrangian approach by including also the spin-$\ffh$ 
resonances. 

The paper is organized as follows. We start in Sec. \ref{model} with 
a description of the formalism  concentrating  mainly on 
the spin-$\ffh$ couplings; the complete discussion of our model including 
all other couplings can be found in \cite{Penner02,pm2,phd}.
In  Sec. \ref{results}  we discuss the results of our calculations in comparison with 
the previous studies \cite{Penner02} and finish with a summary.

\section{The Giessen model}
\label{model}
 We solve the Bethe-Salpeter coupled-channel equation in the $K$-matrix approximation  
to extract scattering amplitudes for the final states under consideration. 
The validity of the  $K$-matrix approximation has been tested by  Pearce and Jennings who 
 have performed a fit to the elastic $\pi N$ phase shifts  
up to 1.38 GeV with the 'smooth',  Blankenbecler-Sugar 
and the K-matrix propagators \cite{pearce}. They have found no significant differences in 
the parameters extracted in the three cases. Also, a successful description 
of the  pion- and photon-induced reaction data  \cite{Penner02,pm2} and  $\eta$-production 
\cite{sauermann} points to  the applicability of this approximation for investigation of the baryon 
resonance spectra. 

In order to decouple
the equations we perform a partial-wave decomposition of the $T$ matrix into
total spin $J$, isospin $I$, and parity $P=(-1)^{J\pm\frac{1}{2}}$. Then the partial-wave amplitudes 
can be expressed in terms of an interaction potential $\mck$ via the matrix equation
\bea
\mct^{I,J\pm} = 
\left[ \frac{\mck^{I,J\pm} }{1 - \mi \mck^{I,J\pm}}\right], \;
\label{bsematinv}
\eea
where each element of the matrices $\mct^{I,J\pm}_{fi}$ and $\mck^{I,J\pm}_{fi}$ 
corresponds to a given initial and final state ($i,f =$ $\pi N$, $2\pi N$, $\eta N$, 
$K \Lambda$, $K \Sigma$, $\omega N$ ). The interaction potential is approximated 
by tree-level Feynman diagrams which in turn are obtained from effective 
Lagrangians \cite{Penner02,phd}. The $\mct$-matrix \refe{bsematinv}  fulfils  
unitarity  as long as the $\mck$ matrix is hermitian.
In our model the following 19
resonances are included
$P_{33}(1232)$,  $P_{11}(1440)$, $D_{13}(1520)$, $S_{11}(1535)$, $P_{33}(1600)$, $S_{31}(1620)$, 
$S_{11}(1650)$,  $D_{15}(1675)$, $F_{15}(1680)$, 
$D_{33}(1700)$,  $P_{11}(1710)$, $P_{13}(1720)$, $P_{31}(1750)$, 
$P_{13}(1900)$, ~~~~~~ $P_{33}(1920)$, $F_{35}(1905)$,
$D_{35}(1930)$, $F_{15}(2000)$,
and $D_{13}(1950)$, which is denoted as $D_{13}(2080)$ by 
the PDG \cite{pdg}.  

The Lagrangian for the spin-$\ffh$ resonance decay to a final baryon $B$ and a (pseudo)scalar  
meson 
$\varphi$ is chosen in the form
\bea
\mcl^{\ffh}_{\varphi B R} = \frac{{\rm g}_{\varphi B R}}{m_\pi^2}\bar u_R^{\mu \nu}
\Theta_{\mu\delta}(a)  \Theta_{\nu\lambda}(a')\Gamma_S   u_B  \partial^\delta \partial^\lambda 
\varphi + h.c.
\label{Lagran52}
\eea
with the  matrix $\Gamma_S=\umat$ if resonance and final meson have identical parity  and  
$\Gamma_S=\mi\gamma_5$  otherwise.
The off-shell projector $\Theta_{\mu\nu}(a)$
is defined by 
\bea
\Theta_{\mu\nu}(a) = {\rm g}_{\mu\nu} - a\gamma_\mu\gamma_\nu,
\label{off}
\eea
where $a$ is a free off-shell parameter. Since the on-shell  symmetric spin-$\ffh$  field $u_R^{\mu \nu}$ 
has to obey the Dirac equation and satisfies the conditions 
$\gamma_\mu u_R^{\mu \nu}=\partial_\mu u_R^{\mu \nu} ={\rm g}_{\mu\nu} u_R^{\mu \nu} =0$ \cite{rarita}
the second  part in \refe{off} only contributes for off-shell particles,
giving rise to  lower off-shell spin components in \refe{Lagran52}. 
In general the interaction Lagrangian \refe{Lagran52} can have two off-shell projectors matched with 
both vector indices of the resonance field tensor. However, as we will see later, a good description 
of the experimental data can be achieved already with a single 
parameter $a$ keeping the second one equal to zero. Thus, to keep our model as simple as possible 
we use only  one off-shell projector in \refe{Lagran52}. 

The widths of the hadronic resonance decays as extracted from the Lagrangian  
\refe{Lagran52}
are 
\bea
\Gamma_\pm(R_\ffh\to\varphi B)= I\frac{{\rm g^2_{\varphi B R}}}{30\pi m_\pi^4}k_\varphi^5
\frac{E_B\mp m_B}{\sqrt{s}}.
\label{width}
\eea
The upper sign corresponds to the decay of a resonance into a meson with the  
identical parity and
vice versa. $I$ is the isospin factor and $k_\varphi$, $E_B$, and $m_B$ are the meson  momentum, 
energy and mass of the final baryon, respectively. 

The coupling  of the spin-$\ffh$ resonances to the   
$\omega N$ final state is chosen  to be 
\bea
\mcl^{\ffh}_{\omega N} =
\bar u_R^{\mu \lambda}\Gamma_V
\left( \frac{{\rm g}_1}{4m_N^2}\gamma^\xi
+\mi\frac{{\rm g}_2}{8m_N^3}\partial^\xi_{N}
+\mi\frac{{\rm g}_3}{8m_N^3}\partial^\xi_{\omega}\right)\nonumber\\
\times(\partial^{\omega}_\xi{\rm g}_{\mu\nu} - 
\partial^{\omega}_\mu{\rm g}_{\xi\nu})u_N \partial_\lambda^\omega  \omega^\nu + h.c.,
\label{LgrOm}
\eea
where the matrix $\Gamma_V$ is $\umat$ ($\mi\gamma_5$) for positive (negative) resonance parity 
and $\partial^\mu_N$ ($\partial_\mu^\omega$)  denotes the partial derivative of
the nucleon  and the $\omega$-meson  fields, respectively. 
The above Lagrangian is constructed in 
the same manner as the one for  spin-$\fth$ in \cite{Penner02}. Similar couplings were also used to 
describe 
electromagnetic processes \cite{Titov,Wolf,David,Hand}.   
Since the different parts of \refe{LgrOm} contribute at different kinematical conditions we 
keep all three couplings as  free parameters  and vary them during the fit. 
The helicity amplitudes for the decay $R\to \omega N$ are given by
\bea
A^{\omega N}_{\fth} &=&
\frac{\sqrt{E_N\pm m_N}}{\sqrt{5m_N}}
\frac{k_\omega}{4m_N^2}
(
-{\rm g}_1(m_N\mp m_R)\nonumber\\
&+&{\rm g}_2\frac{(m_R E_N-m_N^2)}{2m_N}
+{\rm g}_3\frac{m_\omega^2}{2m_N^2} ) ,\nonumber\\
A^{\omega N}_{\foh} &=&
\frac{\sqrt{E_N\pm m_N}}{\sqrt{10m_N}}
\frac{k_\omega}{4m_N^2}
( 
{\rm g}_1 (m_N\pm (m_R-2E_N))\nonumber\\ 
&+&{\rm g}_2\frac{(m_R E_N -m_N^2)}{2m_N}
+{\rm g}_3\frac{m_\omega^2}{2m_N^2} ),  \nonumber\\
A^{\omega N}_{0}&=&\frac{\sqrt{(E_N\pm m_N)}}{\sqrt{5m_N}}
\frac{k_\omega m_\omega}{4m_N^2} 
( {\rm g}_1
\pm {\rm g}_2\frac{E_N}{2m_N}\nonumber\\
&\pm& {\rm g}_3\frac{(m_R-E_N)}{2m_N} ),
\label{helic}
\eea
with upper (lower) signs corresponding to positive (negative) resonance parity. The lower 
indices stand for the helicity $\lambda$ of the final $\omega N$ state
$\lambda=\lambda_V -\lambda_N$, where 
we use  an
abbreviation as follows: $\lambda=$ $0: 0+\foh$, ~$\foh:1-\foh$,
~ $\fth:1+\foh$. 
The resonance $\omega N$ 
decay width $\Gamma^{\omega N}$ can be written as the sum over the three helicity amplitudes given above:
\bea
\Gamma^{\omega N}= \frac{2}{(2J+1)}\frac{k_\omega m_N}{2\pi m_R}
\sum_{\lambda=0}^{3/2} |A^{\omega N}_\lambda|^2,
\label{helicW}
\eea
where  $J=\ffh$ for the spin-$\ffh$ resonance decay. 

For practical calculations we adopt the spin-$\ffh$ projector in the form 
\bea
P_{\ffh}^{\mu\nu,\rho \sigma}(q)&=&\frac{1}{2}(
 T^{\mu\rho}T^{\nu \sigma}
+T^{\mu \sigma}T^{\nu \rho} )
-\frac{1}{5}T^{\mu\nu}T^{\rho \sigma}\nonumber\\
&+&\frac{1}{10}(
T^{\mu\lambda} \gamma_\lambda\gamma_\delta T^{\delta \rho}T^{\nu \sigma}
+T^{\nu\lambda} \gamma_\lambda\gamma_\delta T^{\delta \sigma}T^{\mu \rho}\nonumber\\
&& +T^{\mu\lambda} \gamma_\lambda\gamma_\delta T^{\delta \sigma}T^{\nu \rho}
+T^{\nu\lambda} \gamma_\lambda\gamma_\delta T^{\delta \rho}T^{\mu \sigma}),
\label{ConvPr}
\eea
with
\bea
T^{\mu\nu}&=&-{\rm g}^{\mu\nu}+\frac{q^\mu q^\nu}{m_R^2},
\label{ConvPr2}
\eea
which has also  been  used in an analysis of  $K\Lambda$
photoproduction \cite{David}. 

As is well known the description of particles with spin  
$J>\foh$ leads to a number of different propagators which have 
non-zero  off-shell lower-spin  components. To control these components the off-shell 
projectors \refe{off} are usually introduced. There were attempts to fix the off-shell
parameters and remove the spin-$\foh$ contribution in the 
case of spin-$\fth$ particles \cite{nath}.
However, it has been shown \cite{ben89} that these contributions cannot be suppressed for any 
value of $a$. Indeed, Read \cite{Read} has demonstrated that the choice of the off-shell
parameter in the coupling is closely linked to the off-shell behavior of the propagator.
To overcome this problem Pascalutsa suggested
gauge invariance as an 
additional constraint to fix the interaction Lagrangians for higher spins and remove
the lower-spin components \cite{pascatim}. 
Constructing the spin-$\fth$ interaction for a Rarita-Schwinger field $u^\mu_{\fth}$
by only  allowing couplings to the gauge-invariant field 
tensor  $U^{\mu\nu}_{\fth}=\partial^\mu u^\nu_{\fth}-\partial^\nu u^\mu_{\fth}$  Pascalutsa  
derived an interaction  which (for example) for the $\pi N \Delta$ coupling is
\bea
\mcl_{\pi N\Delta}=  f_\pi \bar u_N\gamma_5 \gamma_\mu \widetilde{U}^{\mu\nu}
\partial_\nu \varphi + h.c.,
\label{Psc}
\eea
where $\widetilde{U}^{\mu\nu}$ is the tensor dual to $U^{\mu\nu}$: 
$\widetilde{U}^{\mu\nu} = \varepsilon^{\mu\nu\lambda\rho}U_{\lambda\rho}$ and 
$\varepsilon^{\mu\nu\lambda\rho}$ is the Levi-Civita tensor. 
The same arguments can also be applied to spin-$\ffh$ particles. In this case the
amplitude of meson-baryon scattering can be obtained from the  conventional amplitude 
by the  replacement 
\bea
\Gamma_{\mu\nu}(p',k')\frac{P_{\ffh}^{\mu\nu,\rho \sigma}(q)}{\slash q-m_R}
&&\Gamma_{\rho\sigma}(p,k)\nonumber\\
\to\Gamma_{\mu\nu}(p',k')&&\frac{{\mathcal P}_{\ffh}^{\mu\nu,\rho \sigma}(q)}{\slash q-m_R}
\Gamma_{\rho\sigma}(p,k)\frac{q^4}{m_R^4},
\label{PscCoupl}
\eea
where  $\Gamma_{\rho\sigma}(p,k)$ are vertex functions that follow from 
\refe{Lagran52} and \refe{LgrOm} by applying  Feynman rules
 and the projector ${\mathcal P}_{\ffh}^{\mu\nu,\rho \sigma}(q)$
is obtained from (\ref{ConvPr}, \ref{ConvPr2}) by the replacement 
$$q^\mu q^\nu/m_R^2 \to q^\mu q^\nu/q^2.$$
This procedure is similar to that which has been used in the spin-$\fth$ case 
\cite{pascatim}.  
It has been shown  for the spin-$\fth$ case  \cite{pasca01} ,  that both prescriptions are 
equivalent in the effective Lagrangian approach as long as additional contact interactions
are taken into account when the Pascalutsa couplings are used.  
The differences between these descriptions  have been discussed in 
\cite{Penner02,pascatjon,lahiff} and here we perform calculations by using both 'conventional' ($C$)
and Pascaluta ($P$) approaches. Similar to the spin-$\fth$ case \cite{Read}, the off-shell 
parameters $a$ in (\ref{off}) can be linked to the coupling strengths extracted through (\ref{helic}) 

In order to take into account the internal structure of mesons and baryons each  vertex is 
dressed by a corresponding formfactor:
\bea
F_p (q^2,m^2) &=& \frac{\Lambda^4}{\Lambda^4 +(q^2-m^2)^2}. 
\label{formfacr} 
\eea
Here  $q$ is the four momentum of the intermediate particle and $\Lambda$ is a cutoff parameter.
In \cite{Penner02} it  has been shown that the
 formfactor \refe{formfacr} gives  systematically better 
results  as compared to other ones, therefore we do not use any other forms for $F(q^2)$. The 
cutoffs $\Lambda$ in \refe{formfacr} are treated as  free parameters and allowed to be varied 
during the  calculation. However we demand the same cutoffs in all channels for a 
given  resonance
spin  $J$ : $\Lambda^{J}_{\pi N}=\Lambda^{J}_{\pi\pi N}=\Lambda^{J}_{\eta N}=...$ etc., 
($J=\foh,~ \fth,~ \ffh$). This greatly reduces the number of free parameters; i.e. for all 
spin-$\ffh$ resonances there is  only one cutoff $\Lambda_{\ffh}$ for all decay channels.

To take into account  contributions of the $2\pi N$ channel in our calculations
we use the inelastic partial-wave cross section $\sigma_{2\pi N}^{JI}$ data 
extracted  in \cite{manley84}. To this end the inelastic $2\pi N$ channel is parameterized by 
an effective $\zeta N$ channel where $\zeta$ is an effective isovector meson with mass 
$m_\zeta=2m_\pi$.
Thus  $\zeta N$ is considered as a sum of different
($\pi \Delta$, $\rho N$, etc.) contributions to the total $2\pi N$ flux. 
We allow only resonance $\zeta N$-couplings since  each background diagram would introduce
a meaningless coupling parameter. Despite  this approximation the studies 
\cite{feusti98,feusti99,sauermann,Penner02} have achieved a good description of the total 
partial-wave cross sections \cite{manley84} and we proceed in our calculations by using the above 
prescription. For the $R \ra \zeta N$ interaction  the same Lagrangians are used as for
the $R \ra \pi N$ couplings  taking into account the positive parity of the $\zeta$ meson.

\section{Results and discussion}
\label{results}
We use the same database as 
in \cite{Penner02} with additional elastic $\pi N$ data for the spin-$\ffh$ partial wave 
amplitudes taken from the VPI group analysis \cite{SM00}. 
For the $2\pi N$ channel we use the  spin-$\ffh$ partial  wave cross sections derived in 
\cite{manley84}.  We confine ourselves to the energy region 
$m_\pi + m_N \leqslant \sqrt{s} \leqslant 2$ GeV. 
The database on 
the $\eta N$, $K \Lambda$, $K \Sigma$ and $\omega N$ channels incorporates all available 
 experimental information from the pion threshold up to the 2 GeV energy region. 
This  includes partial and differential cross sections and polarisation measurements. 
The references on these reactions, 34 in total, are summarized 
in \cite{phd} 

The results presented in the following are 
from ongoing calculations to describe the data in all channels simultaneously. 
The resulting $\chi^2$ of our best overall hadronic fits are given in Table \ref{sqr}.
 The obtained $\chi^2_{\pi \pi}$ and $\chi^2_{\pi 2\pi}$ are 
calculated using experimental data  from all $\pi N$ and $2\pi N$  partial waves
up to spin-$\ffh$ except the $D_{35}$ wave. 
We find a problem with the description of the $D_{35}$ partial wave so  
the resulting $\chi^2_{\pi \pi}$
turns out to be very large. 
Hence the  $\chi^2_{\pi \pi}$ values given in Table \ref{sqr}   are  calculated by neglecting  the
$\pi N$ datapoints for the $D_{35}$ partial wave.
From Table \ref{sqr} one can conclude
that the conventional $C$-prescription leads to a better description of the data  in all partial waves. Note,
that since the  $P$ coupling does not have 'off-shell' background we also include additional  
$D_{13}(1700)$ and $S_{31}(1900)$ resonances in the $P$-calculations \cite{Penner02,pm2}.

\begin{figure*}
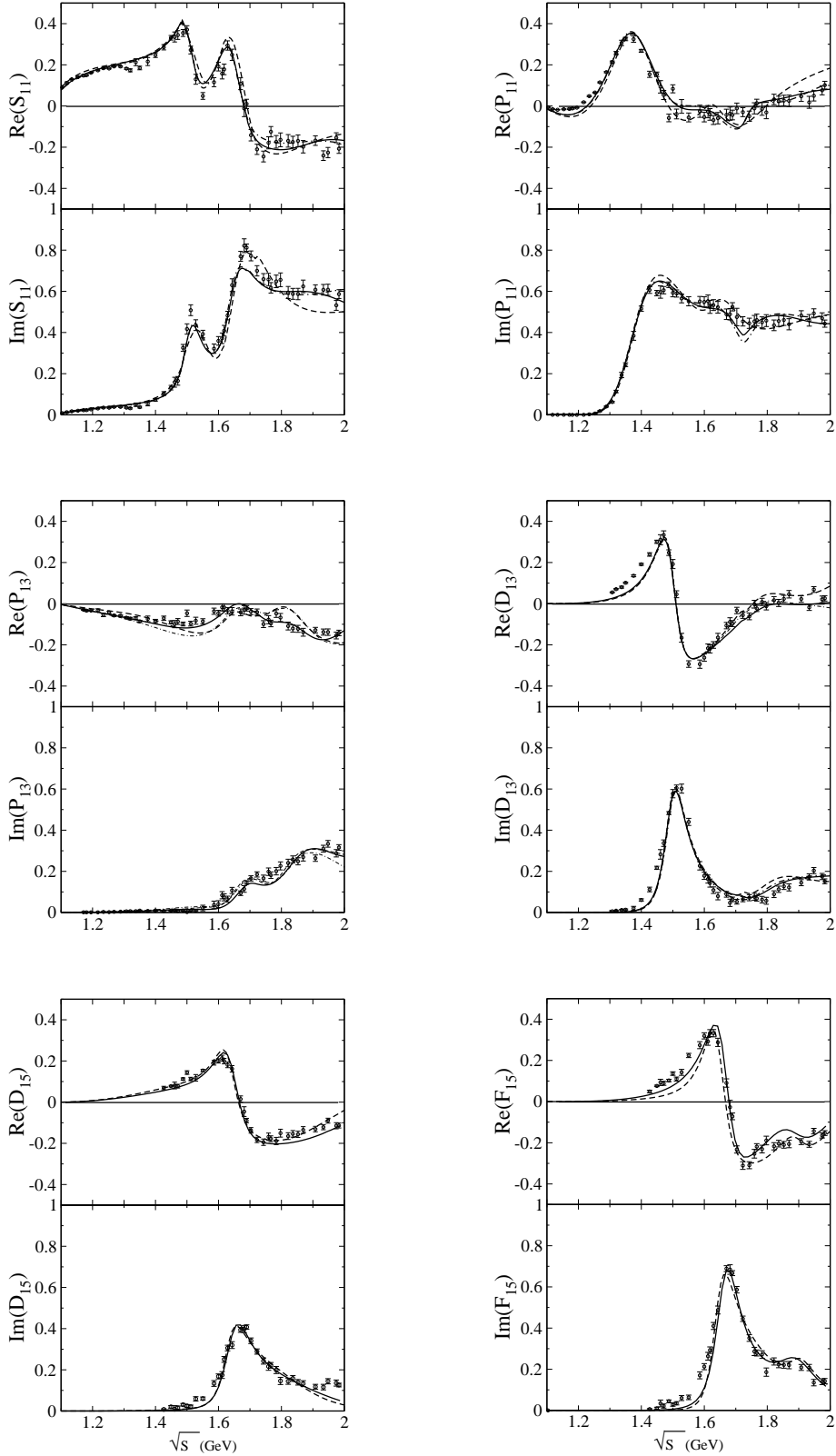

  \begin{center}
    \parbox{16cm}{
\vspace*{5mm}
      \hspace*{1cm}\parbox{70mm}{\includegraphics[width=50mm]{s11n_1.eps}}
                   \parbox{70mm}{\includegraphics[width=50mm]{p11n_2.eps}}\\

\vspace*{9mm}
      \hspace*{1cm}\parbox{70mm}{\includegraphics[width=50mm]{p13n_3.eps}}
                   \parbox{70mm}{\includegraphics[width=50mm]{d13n_4.eps}}\\

\vspace*{9mm}
      \hspace*{1cm}\parbox{70mm}{\includegraphics[width=50mm]{d15n_5.eps}}
                   \parbox{70mm}{\includegraphics[width=50mm]{f15n_6.eps}}\\
       }
       \caption{ 
The $\pi N \to \pi N$ partial waves for $I$=$\foh$.
The solid (dashed) line  corresponds $C$($P$)-calculations. The dash-dotted line
is the best hadronic fit C-p-$\pi +$ from \cite{Penner02}. 
The data are taken from \cite{SM00}. 
      \label{I12}} 
\end{center}
\end{figure*}

\begin{figure*}
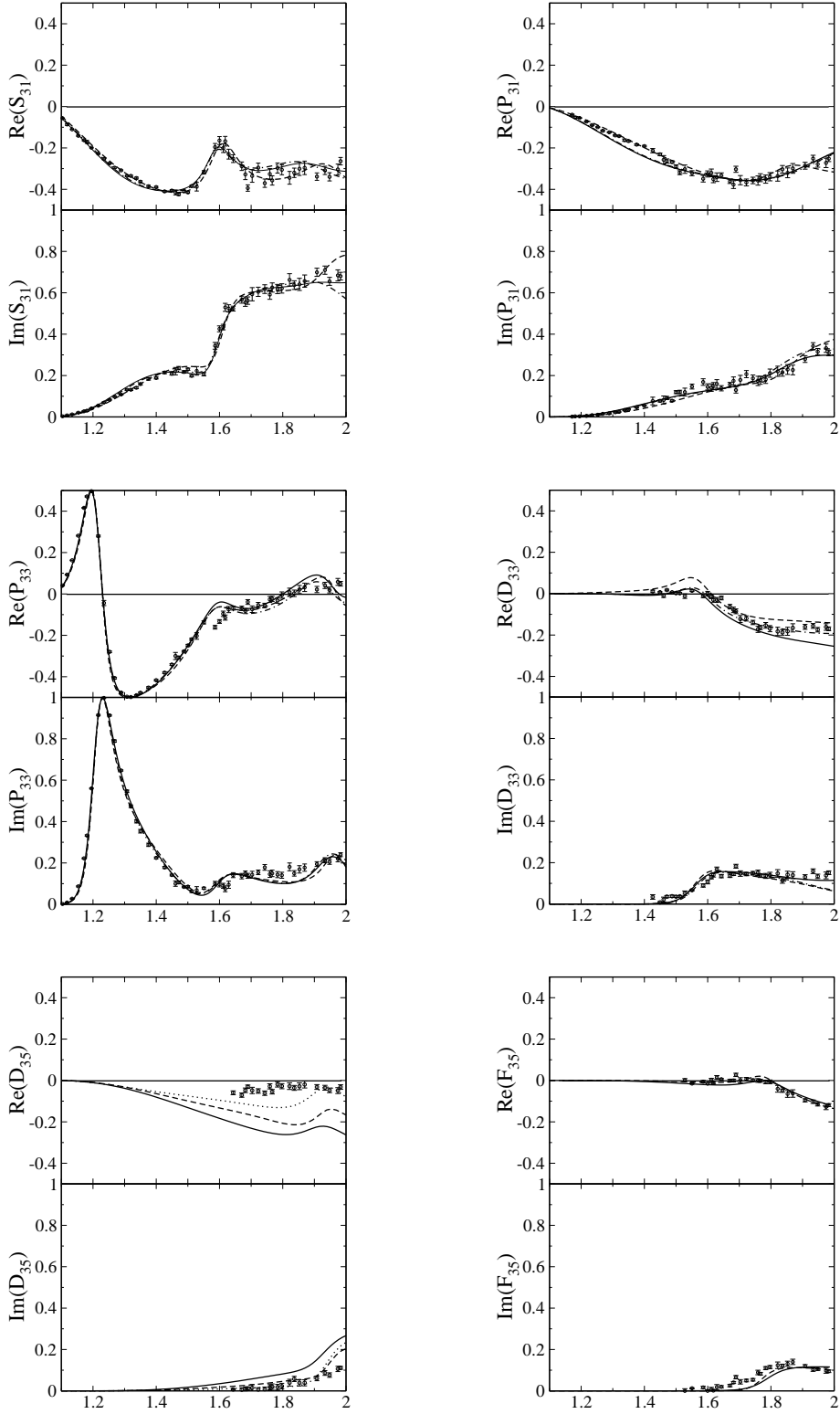

  \begin{center}
    \parbox{16cm}{
\vspace*{5mm}

 \hspace*{1cm}\parbox{70mm}{\includegraphics[width=50mm]{s31n_7.eps}}
              \parbox{70mm}{\includegraphics[width=50mm]{p31n_8.eps}}\\

\vspace{7mm}
  \hspace*{1cm}\parbox{70mm}{\includegraphics[width=50mm]{p33n_9.eps}}
               \parbox{70mm}{\includegraphics[width=50mm]{d33n_10.eps}}\\

\vspace{7mm}
  \hspace*{1cm}\parbox{70mm}{\includegraphics[width=50mm]{d35n_11.eps}}
               \parbox{70mm}{\includegraphics[width=50mm]{f35n_12.eps}}\\
       }
       \caption{ 
The $\pi N \to \pi N$ partial waves for $I$=$\fth$.
The solid (dashed) line  corresponds $C$($P$)-calculations. The dash-dotted line
is the best hadronic fit C-p-$\pi +$ from \cite{Penner02}.  The dotted line is the result
for the $D_{35}$-wave obtained with reduced nucleon cutoff (see text). 
The data are taken from \cite{SM00}. 
      \label{I32}} 
\end{center}
\end{figure*}

Compared to the previous best hadronic fits   C-p-$\pi +$  and    P-p-$\pi +$ from \cite{Penner02,pm2},
we obtain the same values for cutoffs and non-resonant couplings. The only  exception is the 
$\Lambda_{\foh}$=2.79 for the $C$-coupling which is  less than that of  C-p-$\pi +$. In
addition we find 
$g_{NN\omega}$=4.59(4.20) and $\kappa_{NN\omega}$=-0.12(0.06) for $C$($P$) coupling calculations
which slightly differ from \cite{Penner02}. 
\begin{table}
  \begin{center}
    \begin{tabular}
      {l|r|r|r|r|r|r|r}
      \hhline{========}
       Fit & Total $\pi$ & $\chi^2_{\pi \pi}$ & $\chi^2_{\pi 2\pi}$ & 
       $\chi^2_{\pi \eta}$ & $\chi^2_{\pi \Lambda}$ & $\chi^2_{\pi \Sigma}$
       & $\chi^2_{\pi \omega}$ \\ 
       \hline
       $C$  & 2.60 & 2.60 & 7.63 & 1.37 & 2.14 & 1.83 & 1.23 \\
       $P$  & 3.65 & 3.80 &10.06 & 1.75 & 2.54 & 2.93 & 1.83\\
       \hline
    \end{tabular}
  \end{center}
  \caption{$\chi^2$ of the $C$ (first line)  and $P$ fits (second line) . 
  The $D_{35}$ $\pi N$ and $2\pi N$ data have not been taken into account (see text). 
    \label{sqr}}
\end{table}
The results for the $\pi N$ partial wave amplitudes  are shown on the 
Figs. \ref{I12} - \ref{I32} in comparison with  C-p-$\pi +$ result from 
\cite{Penner02}. We do not show here the corresponding P-p-$\pi +$ result since it almost
coincides with the  new $P$-calculations. 
The main differences are found for the conventional coupling calculations in comparison 
with the previous study. A substantially better description in the $P_{13}$  partial wave
is due to the additional off-shell background generated by spin-$\ffh$ resonances. 
 The same effect also improves the  description of  the real and imaginary high energy tails
of the $P_{31}$ and $S_{31}$ amplitudes, respectively. The contribution from the 
spin-$\ffh$ resonances can also be seen in the  $D_{33}$ amplitude which is also  
affected by spin-$\ffh$ off-shell  components. This  leads to 
a worsening in the imaginary part of $D_{33}$ above 1.8 GeV,  giving however improvement in the
corresponding real part.

\begin{figure*}
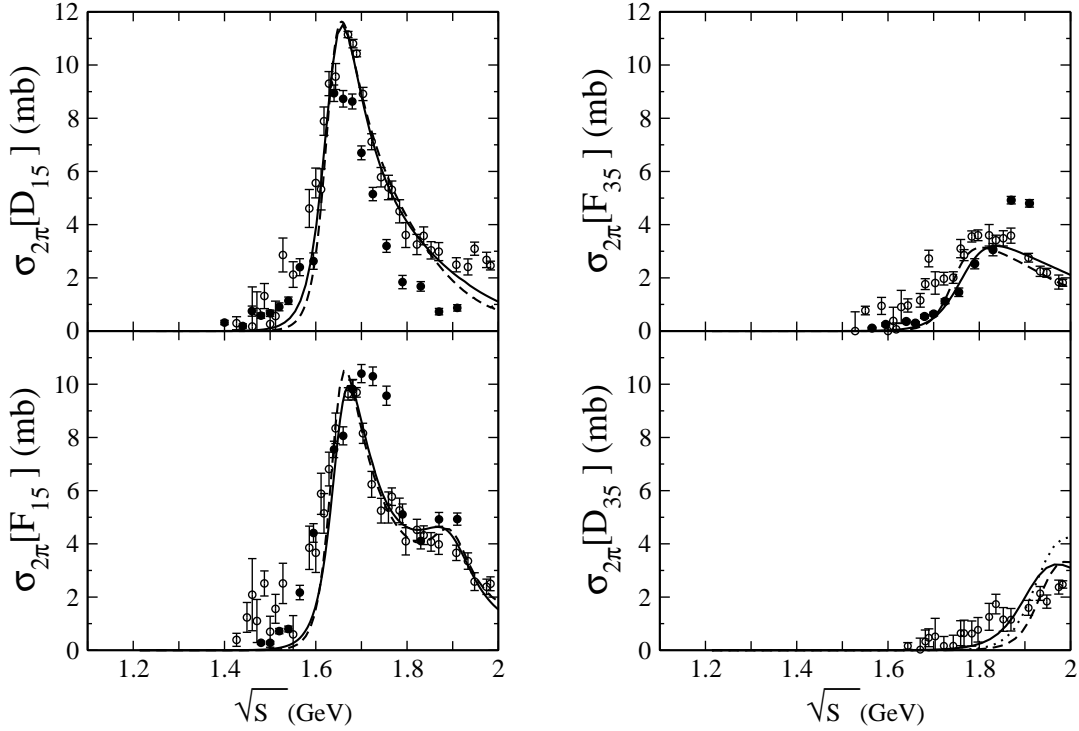

  \begin{center}
    \parbox{16cm}{
      \parbox{75mm}{\includegraphics[width=70mm]{d15f15p2n_13.eps}}
      \parbox{75mm}{\includegraphics[width=70mm]{f35p2n_14.eps}}
       }
       \caption{ 
The inelastic $D_{15}$, $F_{15}$, $F_{35}$, and $D_{35}$ waves. The solid (dashed) line  
corresponds to calculation  $C$ ($P$)  for the $2\pi N$ channel.
Open and filled circles represent
the total inelasticity from the VPI group \cite{SM00} and the $2\pi N$ data \cite{manley84}, 
respectively. The calculated inelasticities 
almost coincide with the calculated $2\pi N$ cross sections  and are  not shown here. 
Calculation with a reduced nucleon cutoff is shown by the dotted line.
      \label{2piN}} 
  \end{center}
\end{figure*}

The  $D_{15}(1675)$,
$F_{15}(1680)$,  and  $F_{35}(1905)$ resonances were included in our 
calculations. We have also found evidence for a second  $F_{15}$ state around 1.98 GeV which 
is rated two-star by \cite{pdg}.
The results for $\pi N\to 2\pi N$ partial wave cross sections are shown in Fig. \ref{2piN}.
We  stress that the $\pi N$ partial  wave inelasticities are not fitted  
but  obtained as a sum of the individual contributions from all open channels.  

In the following each spin-$\ffh$ wave is discussed  separately.  
The extensive discussion of the spin-$\foh$ and spin-$\fth$ partial waves  can 
be found in \cite{Penner02,pm2}. The parameters of the corresponding baryon resonances  
are listed in \cite{www}.
 
{$\bf D_{15}$}. The elastic VPI data show a single resonant peak which corresponds  
to the well established  $D_{15}(1675)$ state. We find a
good description of the elastic amplitude in  both the $C$- and 
$P$-calculations. 

\begin{figure*}
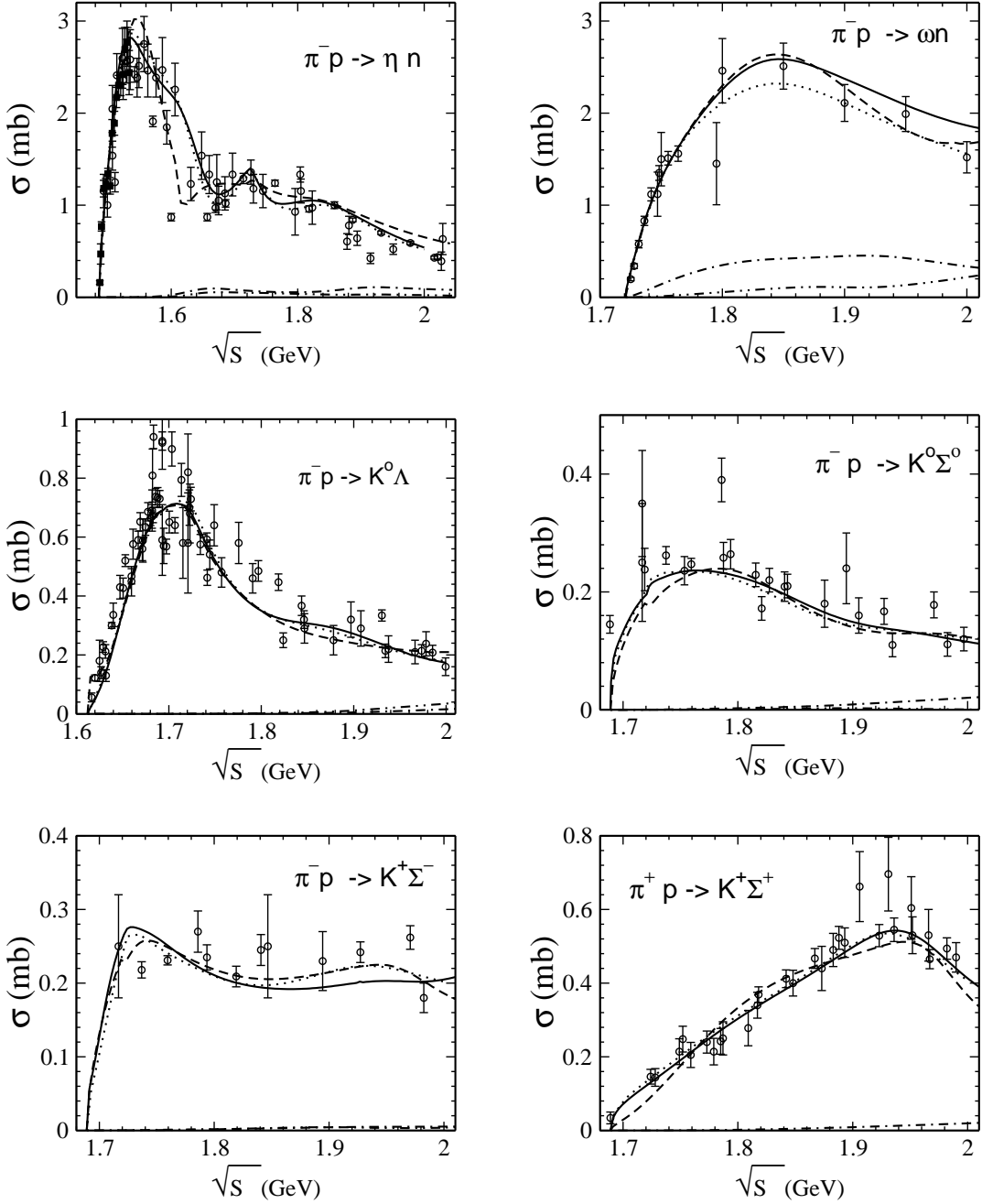

  \begin{center}
    \parbox{16cm}{
\vspace*{5mm}
      \parbox{75mm}{\includegraphics[width=69mm]{eta_tot_15.eps}}
      \parbox{75mm}{\includegraphics[width=69mm]{omega_tot_16.eps}}
       }\vspace*{0.1cm}
    \parbox{16cm}{
      \parbox{75mm}{\includegraphics[width=69mm]{lamb_tot_17.eps}}
      \parbox{75mm}{\includegraphics[width=69mm]{ps0_tot_18.eps}}
       }\vspace*{0.1cm}
    \parbox{16cm}{
      \parbox{75mm}{\includegraphics[width=69mm]{ps2_tot_19.eps}}
      \parbox{75mm}{\includegraphics[width=69mm]{ps3_tot_20.eps}}
       }
       \caption{ 
The total cross sections for the inelastic reactions. 
The solid (dashed) line  corresponds to the  $C$ ($P$)
 result. The dotted line  shows our previous results C-p-$\pi +$ 
\cite{Penner02}. The contributions from the spin-$\ffh$ states are shown by dash-dotted ($C$) 
and dash-double-dotted ($P$) lines. 
For the data references see 
\cite{Penner02}.\label{crstot}} 
  \end{center}
\end{figure*}

The $2\pi N$ data  \cite{manley84} are systematically below the total inelasticity of the VPI 
group \cite{SM00}.
 This can be an indication that apart from  $2\pi N$  there are 
additional contributions from other inelastic channels. 
However, in the analysis of Manley and Saleski 
\cite{manley92} as well as in the most recent 
study of Vrana et al. \cite{vrana}  the total inelasticity in the $D_{15}$ wave is
entirely  explained by the resonance decay to the $\pi \Delta$ channel. We also find no
significant  contributions from the $\eta N$, $K\Lambda$, $K\Sigma$, and $\omega N$ channels
to the total $\pi N$ inelasticity in the present hadronic calculations. 
The calculated $2\pi N$ cross sections are found to be substantially above the data 
from \cite{manley84} in all fits. 
Indeed, the difference between the $2\pi N$  and  inelasticity 
data runs into 2 mb at 1.67 GeV. This flux can be absorbed by neither $\eta N$, $K\Lambda$, 
$K\Sigma$, $\omega N$ channels, see Fig.\,\ref{crstot}. Thus we conclude that either the 
$\pi N$ and $2\pi N$ data are inconsistent with each other or other open channels 
(as 3$\pi N$) must be taken 
into account. To overcome this problem and to describe the $\pi N$ and $2\pi N$ data in 
the $D_{15}$  partial wave the original $2\pi N$ data error bars \cite{manley84} 
 were enlarged by a
factor 3. The same procedure was also used by Vrana et al. \cite{vrana} and Cutkosky et al. 
\cite{cutkosky90} to fit the inelastic data.

In the both $C$ and $P$ coupling calculations the  total inelasticities in the  
$D_{15}$ wave almost coincide with the partial-wave cross sections and therefore are not 
shown in Fig.\,\ref{2piN} (left top). A good description of the inelasticity in the $D_{15}$ 
wave is achieved and the extracted resonance parameters are also in agreement with other 
findings (see next section).

\begin{table*}[t]
  \begin{center}
    \begin{tabular}
      {l|c|r|r|r|r|r|r|r}
      \hhline{=========}
      $L_{2I,2S}$ & mass & $\Gamma_{tot}$ &
      $R_{\pi N}$ & $R_{2\pi N}$ & $R_{\eta N}$ &
      $R_{K \Lambda}$ & $R_{K \Sigma}$ & $R_{\omega N}$ \\
      \hhline{=========}
$D_{15}(1675)$& 1665 & 144 & 40.2 & $59.1(-)$ & $ 0.6(-)$ & $ 0.0(+)$ & $-0.04^a$ & --- \\
              & 1662 & 138 & 41.2 & $58.4(+)$ & $ 0.4(-)$ & $ 0.0(-)$ & $ 0.02^a $ & --- \\
      \hhline{=========}
$F_{15}(1680)$& 1674 & 120 & 68.5 & $31.5(-)$ & $ 0.1(+)$ & $ 0.0(+)$ & $ 0.07^a$ & --- \\
              & 1669 & 122 & 65.8 & $34.2(+)$ & $ 0.0(-)$ & $ 0.0(+)$ & $ 0.13^a$ & --- \\
     \hline
$F_{15}(2000)$& 1981 & 361 &  9.0 & $84.0(+)$ & $ 4.3(-)$ & $  0.5^b(-) $ & $ 0.4(-)$ &  2.2 \\
              & 1986 & 488 &  9.5 & $88.2(-)$ & $ 0.3(-)$ & $ 0.1(+)$ & $ 0.2(-)$ &  1.7 \\
      \hhline{=========}
$F_{35}(1905)$& 1859 & 400 &  11.3 & $ 88.7(+)$ & --- & ---  & $ 0.7^b(+)$ & --- \\
              & 1830 & 457 &  10.3 & $ 89.7(-)$ & --- & ---  & $ 0.0(+)$ & --- \\
      \hhline{=========}
    \end{tabular}
  \end{center}
  \caption{
    Properties of the  spin-$\ffh$ resonances considered in the
    present calculation.
    Masses and total widths $\Gamma_{tot}$ are given 
    in MeV, the decay ratios $R$ in percent of the total width.  
    In brackets, the sign of the coupling is given (all $\pi N$ couplings
    are chosen to be positive).
    $^a$: The coupling is presented since the resonance is
    below threshold. 
    $^b$: Decay ratio in 0.1\permil. 
    The first line corresponds to $C$-calculation  and the second one to 
    $P$.
    \label{52param}} 
\end{table*}

{$\bf F_{15}$}. The  $F_{15}(1680)$ and $F_{15}(2000)$ resonances are identified in 
this partial  wave. 
The inclusion of the second  resonance significantly improves the description of the
$\pi N$ and $2\pi N$ experimental
data in the higher-energy region. Some evidence for this state was  also found in earlier 
works \cite{manley92,hohler}. 
A visible inconsistency  
between the inelastic VPI data and  the 
$2\pi N$ cross section from \cite{manley84} above 1.7 GeV  can  be seen in Fig. \ref{2piN} 
(left bottom). 
The three  data points at 1.7, 1.725, and 1.755 GeV have, therefore, not been included in our 
calculations. 
Finally we achieve a reasonable description for both $\pi N$ and $2\pi N$ data. The 
$C$ and $P$ coupling calculations give approximately the same results.  

{$\bf F_{35}$}. A single resonance state $F_{35}(1905)$ was taken into account.
Some other  models find an additional lower-lying resonance with a mass of about 1.75 GeV
\cite{manley92,cutkosky79,hohler,vrana}. 
However, we already find a good description of the elastic 
$\pi N$ amplitudes and the $2\pi N$ cross sections by only including the single $F_{35}(1905)$ 
state. 
The inclusion of a second state with somewhat lower mass leads to a worse
description of the $\pi N$ and $2\pi N$ data due to the strong interference between 
the two  nearby states. 
The two $2\pi N$ data points at 1.87 and 1.91 GeV,  which are apparently above the total 
inelasticity, have  not been  included in the calculations.   

The total inelasticity in the $F_{35}$ partial wave almost coincides 
with the calculated
$2\pi N$ cross section and is not shown in Fig.\,\ref{2piN}. Note, that the $2\pi N$ data 
at 1.7 GeV are slightly below the total inelasticity from
\cite{SM00}. This could indicate that other inelastic channels (as $3\pi N$) give additional 
contributions to this partial wave.

There are also difficulties in the description of the $2\pi N$ low-energy tails of the $D_{15}$ 
and $F_{15}$ partial waves below 1.6 GeV, where the calculated cross sections
are slightly below the $2\pi N$ data. The discrepancy leads to a significant rise in 
$\chi^2_{\pi 2\pi}$ (cf. Table \ref{sqr}). The same  behavior 
has been found in our previous calculations for the $D_{I3}$ partial waves \cite{Penner02}. 
There, it has been suggested that the problem might be caused by the description of the
$2\pi N$ channel in terms of an effective $\zeta N$ state. Indeed,
the findings of \cite{manley92,vrana} show   strong $\pi \Delta$ decay ratios  in 
all three
$D_{15}$,  $F_{15}$, and  $F_{35}$ partial waves. The description of the $2\pi N$ channel in terms 
of the $\rho N$ and $\pi \Delta$  channels may change the situation when
taking into account the
$\rho N$ and $\pi \Delta$ phase spaces and corresponding spectral functions. 
Upcoming calculations will address this question.

{$\bf D_{35}$}.  A single $D_{35}(1930)$ resonance is taken into account.
However, there is no clear resonance structure in  the $\pi N$ data for this partial wave.
The data \cite{SM00} also show a total inelasticity at the 2 mb 
level whereas the $2\pi N$ channel was found to be negligible \cite{manley84}. 
It has been suggested \cite{manley84} that this channel could have an important  
inelastic $3\pi N$ contribution. 
Since the measured $2\pi N$ cross section is zero we have used the inelastic $\pi N$ data with
enlarged error bars instead of  the $2\pi N$ data to pin down the $2\pi N$ $D_{35}$ contributions. 
Even in this case  we have found difficulties in the description of the $D_{35}$ partial wave. 
The $\pi N$ channel turns out to be strongly influenced by the $u$-channel nucleon and 
resonance contributions which give significant contributions to the real part of $D_{35}$. 
As can be seen in Fig.\,\ref{I32} the $C$- and $P$-coupling calculations 
cannot give even a rough description of the experimental data \cite{SM00}. 
The situation can be improved by either using a reduced nucleon cutoff $\Lambda_N$ 
or by neglecting the nucleon $u$-channel contribution in the interaction kernel. 
The latter approximation has been used  in the coupled-channel approach of Lutz et al. \cite{lutz}. 
To illustrate this point we have carried out an additional fit for the $C$-coupling
with  the reduced cutoff 
$\Lambda_N$=0.91 taking only the $\pi N$ and $2\pi N$ data into account. The calculated  $\chi^2$ are 
$\chi^2_{\pi \pi}$=3.63 and  $\chi^2_{\pi 2\pi}$=7.87 where the $D_{35}$ data are also taken 
into account (note, that all values in Table\,\ref{sqr} are calculated 
by neglecting these datapoints ). The results for the $D_{35}$ partial wave are shown in 
Fig.\,\ref{I32} by the dotted line. In all calculations for $D_{35}$ presented in 
Fig.\,\ref{I32} the $D_{35}(1930)$ mass was found to be  about 2050 MeV. 
One sees that the calculations with  a reduced nucleon cutoff lead to a better 
description of the $D_{35}$ data giving, however, a worse description of other 
$\pi N$ partial-wave data. Note, that a reduction of the nucleon cutoff is required for a  successful 
description of the lower-spin photoproduction multipoles \cite{Penner02,pm2}
which also leads to  a worsening in $\chi^2$ for the $\pi N$ elastic  channel. 

Finally, we  conclude that the main features of the  considered spin-$\ffh$ partial  waves 
except for $D_{35}$  are well reproduced. 
From \mbox{  Figs. \ref{I12}-\ref{2piN}}
one can see that there is no significant difference between the conventional 
\refe{ConvPr} and the Pascalutsa \refe{PscCoupl} spin-$\ffh$ couplings. 

The parameters of the spin-$\ffh$ resonances are presented in Table \ref{52param}. We
note that the total resonance widths  calculated here do not
necessarily  coincide
with the full widths at half maximum because of the energy dependence of
the decay widths (\ref{width}, \ref{helicW}) 
and the  formfactors used \cite{Penner02}. 
 We do not show here the parameters of the $D_{35}(1930)$ resonance because of the problems
in the  $D_{35}(1930)$ partial wave.
 Although a good description 
of the experimental data  is achieved   some differences in  the extracted resonance parameters
for the $C$- and the $P$-couplings calculations exist.

We obtain a little lower mass for the  $D_{15}(1675)$ as compared to that  obtained
by Manley and Saleski \cite{manley92} and 
\mbox{Vrana et al. \cite{vrana}}, 
but in agreement with other findings \cite{batinic,cutkosky80}. 
The total width is found 
to be consistent with the results from \cite{cutkosky80,hohler,vrana}. 
In the $\eta N$ channel our calculations show a small ($\approx$0.6\%) decay fraction 
which is somewhat higher than the value obtained by 
Batini\'c et al.: 0.1$\pm 0.1$ \%  \cite{batinic}, whereas Vrana et al. give another
bound: $\pm 1$\%. We conclude that both fits give approximately the same 
results for the resonance masses and branching ratios.

The properties of the $F_{15}(1680)$ state are found to be in good agreement with the 
values recommended by \cite{pdg}. We find a somewhat smaller branching ratio in the $\eta N$
channel as compared to that of \cite{batinic}. However, the obtained value  $R_{\eta N}=$0.1\%
is again  in  agreement with the findings of Vrana et al.\,\cite{vrana}: $\pm 1$\%.
The parameters of the second $F_{15}(2000)$  resonance 
differ strongly in various analyses:
Manley and Saleski \cite{manley92} give   $490\pm 310$ MeV for the total decay width
while other studies \cite{arndt95,hohler} find 
it at the level of $95-170$ MeV. Moreover, this state has not been identified in the 
investigations of \cite{vrana,batinic}.
Although we find different results for  $\Gamma_{tot}$ in the two independent calculations, 
the branching ratios are close to each other. A small decay 
width of about  4.3\% is found for the $\eta N$ channel ($C$).  
However, since the $F_{15}(2000)$ resonance is found to be strongly inelastic with  
84-88\% of inelasticity absorbed by the $2\pi N$ channel,  more
 $2\pi N$ data above 1.8 GeV (cf. Fig. \ref{2piN}) are needed for a reliable determination 
of the properties of this state.

The parameters of the $F_{35}(1905)$ state are  in good agreement 
with \cite{pdg}. Both
fits give approximately the same result for the decay branching  ratios.  

All considered resonances have  a rather small  decay ratios to the
$\eta N$, $K\Lambda$, $K \Sigma$, and 
$\omega N$ channels. The only exception is the  $F_{15}(2000)$ resonance
where a small decay  width to $\eta N$  has been found for the conventional coupling 
calculations. 

In Fig. \ref{crstot} the results  for 
$\eta N$, $K\Lambda$, $K \Sigma$, and 
$\omega N$ total cross sections are shown in comparison with best hadronic fit C-p-$\pi +$
from \cite{Penner02}.
The main difference from the  previous result is found in the 
$\omega N$ final state where a visible effect from the inclusion of the spin-$\ffh$ resonances  
is found in the $C$-calculations. 
Although the  $D_{15}(1675)$ and $F_{15}(1680)$ states are below
the $\omega$ production threshold, they give noticeable contributions in the 
$C$-coupling calculations. 
This effect is, however, less pronounced 
in the  $P$-calculations where the role of $D_{15}(1675)$ and $F_{15}(1680)$ are found to be less
important.  

Since the hadronic $\omega N$ data include 
about 115 datapoints the couplings to 
the $\omega N$ channel are not well  constrained and inclusion of photoproduction data may
change the situation \cite{Penner02}.
Looking to the $\omega$-photoproduction reaction the new SAPHIR data may give  an opportunity to
distinguish between various reaction mechanisms. 
We are presently working on this \cite{new}.

\section{Summary and outlook}
We have performed a first investigation of the pion-induced reactions on the nucleon
within the effective Lagrangian
coupled-channels approach including  spin-$\ffh$ resonances.
To investigate the  influence of additional background from the 
spin-$\fth$ and -$\ffh$ resonances 
calculations using both the conventional and the  Pascalutsa higher-spin couplings have been carried
out. A good description of the available  experimental  data has been achieved in all 
$\pi N$, $2\pi N$, $\eta N$, $K \Lambda$, $K\Sigma$, and $\omega N$ final states within both  
frameworks. The $\chi^2$ is somewhat worse for the Pascalutsa prescription, but this is
at least partly due to the absence of additional off-shell parameters in these couplings. 
In view of this ambiguity in the coupling it is gratifying to see that both coupling schemes 
lead to similar physical results for the baryon properties. The effective Lagrangian 
model used in our calculations imposes  stringent physical constraints on the various channels 
and, in particular, on the interplay of the resonance and background contributions. The latter 
are generated by the same Lagrangian  without   any  new unphysical parameters. Thus any remaining 
discrepancy between the data and the calculation points to the necessity to improve our
understanding of the meson-baryon interactions further, for example, by including additional 
t-channel exchanges.

Apart from $2\pi N$ we find no significant contributions  from  other channels to the total 
$\pi N$ inelasticities in the spin-$\ffh$ waves.
Nevertheless, the contributions from higher-spin resonances can be important in  the
$\omega$-production channel. More  data on this reaction are highly desirable to establish the
role of different reaction mechanisms.  

We have  found  evidence for  $F_{15}(2000)$ resonance which is rated two-star by \cite{pdg} 
and has not been included in the most recent resonance analysis by Vrana et al. \cite{vrana}.
However, more precise $\pi N$ and $2\pi N$ data are necessary to 
to identify this state more reliably in  purely hadronic calculations.

For a complete description of $\pi N$ scattering  up to higher energies the J=$\ffh$ resonances 
are obviously needed. 
Compared to our previous study we arrive at a better description in the  $\pi N$, $\eta N$, and
$\pi \Sigma$ channels for the conventional coupling calculations.
Looking only at the lower partial-waves, the improvement in $\pi N$  
is only possible  due to  the additional off-shell background from the spin-$\ffh$ resonances.
On the other hand, the missing background in the Pascalutsa prescription is compensated by 
contributions  from the  $D_{15}$ and $F_{15}$ resonances allowing for a better description 
in the $\eta N$ and $\omega N$ final states.

We are proceeding with the extension of our model by performing a combined analysis of
pion- and photon-induced reactions taking into account spin-$\ffh$ states.
Moreover, the decomposition of the $2\pi N$ channel into  $\rho N$, 
$\pi \Delta$ etc. states will be the subject of further investigations.

\end{document}